\def\e{{\rm e}}
\def\del{\partial}
\def\half{{1\over2}}

\def\abs#1{{\left|{#1}\right|}}
\def\vev#1{\langle #1 \rangle}
\def\del{\partial}
\def\half{{1\over2}}

\def\abs#1{{\left|{#1}\right|}}
\def\vev#1{\langle #1 \rangle}
\def\del{\partial}
\def\dslash{\del\kern-0.55em\raise 0.14ex\hbox{/}}

\def\rough#1{\raise.3ex\hbox{$#1$\kern-.75em\lower1ex\hbox{$\sim$}}}
\newcommand{\PRD}[3]{{\it Phys. Rev.} {\bf D{#1}} (19{#3}) {#2}}
\newcommand{\PRDO}[3]{{\it Phys. Rev.} {\bf {#1}} (19{#3}) {#2}}
\newcommand{\PRDM}[3]{{\it Phys. Rev.} {\bf D{#1}} (20{#3}) {#2}}

\newcommand{\PRLM}[3]{{\it Phys. Rev. Lett.} {\bf {#1}} {#2} (20{#3})}
\newcommand{\NPB}[3]{{\it Nucl. Phys.} {\bf B{#1}} {#2} (19{#3})}
\newcommand{\NPBM}[3]{{\it Nucl. Phys.} {\bf B{#1}} (20{#2}) {#3}}
\newcommand{\PLB}[3]{{\it Phys. Lett.} {\bf {#1}B} (19{#3}) {#2}}
\newcommand{\PLBM}[3]{{\it Phys. Lett.} {\bf B{#1}} (20{#3}) {#2}}

\newcommand{\PTP}[3]{{\it Prog. Theor. Phys.} {\bf {#1}} (19{#3}) {#2}}
\newcommand{\PTPM}[3]{{\it Prog. Theor. Phys.} {\bf {#1}} (20{#3}) {#2}}
\newcommand{\ANN}[3]{{\it Ann. Phys. (N.Y.)} {\bf {#1}}, {#2} (19{#3})}

\newcommand{\MPL}[3]{{\it Mod. Phys. Lett.} {\bf A{#1}} (19{#3}) {#2}}
\newcommand{\MPLM}[3]{{\it Mod. Phys. Lett.} {\bf A{#1}} (20{#3}) {#2}}

\newcommand{\jhep}[3]{{\it JHEP} {\bf {#1}} (20{#2}) {#3}}

\newcommand{\hepph}[1]{{\tt hep-ph/#1}}
\newcommand{\hmu}{\hat\mu}

\documentclass[12pt]{article}
\usepackage{graphicx}
\begin{document}
\baselineskip=18pt
%%%%%%%%%%%%%%%%%%%%%%%%%%%%
\begin{titlepage}
%%%%% PREPRINT NUMBERS %%%%%%
\begin{flushright}
%28/04/2005
KOBE-TH-06-04\\
TU-776
%hep-th/0305xxx
\end{flushright}
\vspace{1cm}
%%%%%%%%%%%%%%%%%%% TITLE %%%%%%%%%%%%%%%%%%
\begin{center}{\Large\bf 
Large Gauge Hierarchy in Gauge-Higgs Unification}
\end{center}
%%%%%%%%%%%%%%%% AUTHORS %%%%%%%%%%%%%%%%%%%%%%%
\vspace{1cm}
\begin{center}
Makoto Sakamoto$^{(a)}$
\footnote{E-mail: dragon@kobe-u.ac.jp} and
Kazunori Takenaga$^{(b)}$
\footnote{E-mail: takenaga@tuhep.phys.tohoku.ac.jp}
\end{center}
%%%%%%%%%%%%%%%%%%%%%%% AFFILIATION %%%%%%%%%%%%
\vspace{0.2cm}
\begin{center}
%\small
${}^{(a)}$ {\it Department of Physics, Kobe University, 
Rokkodai, Nada, Kobe 657-8501, Japan}
\\[0.2cm]
${}^{(b)}$ {\it Department of Physics, Tohoku University, 
Sendai 980-8578, Japan}
%%%%%
%%%%%%%
\end{center}
%%%%%%%%%%%%%%%%%% ABSTRACT %%%%%%%%%%%%%%%
\vspace{1cm}
\begin{abstract}
We study a five dimensional $SU(3)$ nonsupersymmetric 
gauge theory compactified 
on $M^4\times S^1/Z_2$ and discuss the gauge hierarchy 
in the scenario of the gauge-Higgs unification. Making use of
calculability of the Higgs potential and a curious feature that 
coefficients in the potential are given by discrete 
values, we find two models, in which the large gauge 
hierarchy is realized, that is, the weak scale is naturally 
obtained from an unique large scale such as a grand unified
theory scale or the Planck scale. The size of the Higgs mass 
is also discussed in each model. One of the models we find 
realizes both large gauge hierarchy and consistent
Higgs mass, and shows that the Higgs mass becomes heavier as the
compactified scale becomes smaller.
\end{abstract}
\end{titlepage}
%%%%%%%%%%
%\tableofcontents
%%%%%%%%%%%%
\newpage
%%%%%%%%%%%%%%%% INTRODUCTION %%%%%%%%%%%%%%%
\section{Introduction}
The standard model has had a great success in the last 
three decades, and at the moment, there is no discrepancy
of the prediction from precision measurements. The standard 
model, however, has potential shortcomings, one of which 
is stability of the Higgs sector against radiative 
corrections. Namely, the Higgs mass is sensitive to 
ultraviolet effects because of quadratic dependence on
cutoff. Two tremendously separated energy scales cannot be 
stabilized without fine tuning of parameters, and 
there is the gauge hierarchy problem in the standard model. 
\par
%%%%%%%%%%
The problem entirely comes from lack of symmetry to 
control the Higgs sector. Any attempts to overcome 
the problem always lead us to physics beyond the 
standard model. Supersymmetry (SUSY) has been 
considered as one of the promising 
candidates. It controls the Higgs sector to 
suppress the ultraviolet effect on 
the Higgs mass. The gauge hierarchy problem 
is (technically) solved by introducing
SUSY. However, SUSY does not possess 
the mechanism to yield the large gauge 
hierarchy naturally at the tree-level.
The solution to the gauge hierarchy
problem requires the mechanism to generate the 
large gauge hierarchy at the tree-level and 
at the same time, it must be stable against 
radiative corrections. SUSY guarantees only 
the stability against radiative corrections. 
Moreover, the superpartners, which are the 
prediction of supersymmetry, have not yet been 
discovered. Hence, it is important to seek a different 
approach to the gauge hierarchy problem.
\par
%%%%%%%%%
Recently, higher dimensional gauge symmetry 
has been paid much attention as a new approach 
to the gauge hierarchy problem without resorting 
to supersymmetry. The higher dimensional gauge 
symmetry plays the role to control the Higgs 
sector. In particular, the gauge-Higgs 
unification, originally proposed by 
Manton \cite{manton} and Fairlie \cite{fair}, is 
a very attractive idea of physics beyond the 
standard model \cite{gaugehiggs1}. Four dimensional 
gauge and Higgs fields are unified into a 
higher dimensional gauge field, and the theory 
is completely controlled by the higher 
dimensional gauge symmetry. Hence, the mass term for 
the Higgs field is forbidden by the gauge invariance. 
The gauge-Higgs unification has been studied 
extensively from various points 
of view \cite{gaugehiggs2}\cite{models}. 
\par
%%%%%%%%%
In the gauge-Higgs unification, the Higgs field 
corresponds to the Wilson line phase, which
is nonlocal quantity. The Higgs potential is 
generated at quantum level after the compactification 
and, reflecting the nonlocality of the Higgs 
field, the potential never suffers from the ultraviolet 
effect \cite{masiero}. As a result, the Higgs mass 
calculated from the potential is finite as 
well. In other words, the Higgs mass and the 
potential are calculable. This is a very remarkable 
fact, which rarely happens in the usual quantum field 
theory. The specific feature is entirely due to 
shift symmetry manifest through the Wilson line 
phase (Higgs field), which is a remnant 
of the higher dimensional gauge symmetry. Thanks 
to the finite Higgs mass, two tremendously 
separated energy scales can be stable in 
the gauge-Higgs unification.
\par
%%%%%%%%%
The compactification scale is set by the magnitude of the 
vacuum expectation values (VEV) of the Wilson line phase in 
the gauge-Higgs unification when the extra dimension is flat
\footnote{The gauge-Higggs unification with a warped extra 
dimension has been studied in \cite{warp}}. In the 
usual scenario, the scale is at around a few 
TeV \footnote{This is the reason that the gauge-Higgs 
unification can be used to solve the little hierarchy 
problem.}. In this paper, we discuss whether or not 
the compactification scale can be an enormously 
large scale such as a GUT scale or the Planck 
scale \footnote{A different attempt to solve the large gauge
hierarchy problem has been made in the Higgsless gauge symmetry
breaking scenario \cite{nagasaka}.}. 
%%%%%%%%%%%
In discussing the gauge hierarchy, we make use of 
a curious feature of the Higgs potential 
in the gauge-Higgs unification. 
The dynamics of the potential is mostly
governed by massless bulk matter introduced into the theory. 
For small VEV of the Higgs field, in which we
are really interested, the Higgs potential is 
approximated in terms of the logarithm and polynomials with their
coefficients being discrete values given by the flavor number. This 
curious feature is hardly observed in 
the usual quantum field theory, in which the
coefficients are the continuous, scale-dependent parameters. 
\par
%%%%%%%%%%%%%%%
The point for obtaining the large gauge hierarchy
is that the coefficient for the mass
term can be set to zero. We introduce massless bulk matter
satisfying not only the periodic boundary condition
for the $S^1$ direction but also the antiperiodic 
one. The massless bulk field satisfying the antiperiodic 
boundary condition has an opposite sign for the 
coefficient of the mass term from that 
satisfying the periodic one. This is why the 
coefficient can be set to zero even though we do 
not introduce supersymmetry. We also notice that 
this is not the fine tuning of the parameter, but 
just a choice of the flavor set because the 
coefficient is the discrete value given by the 
flavor number.
\par
%%%%%%%%%%%%%
We consider a five dimensional $SU(3)$ gauge 
theory, where one of spatial coordinates is 
compactified on an orbifold $S^1/Z_2$. We
find two models, in which the large gauge 
hierarchy is realized. In model I, the form of 
the Higgs potential is reduced to 
the massless scalar field theory in the
Coleman-Weinberg's paper \cite{cw}. The large gauge 
hierarchy is interpreted as the magnitude 
of the ratio between the logarithmic and 
quartic terms in the Higgs potential. The 
flavor number of the massless bulk matter 
directly affects the size of the gauge 
hierarchy in the model I.
\par
%%%%%%%%%%%
In model II, we introduce massive bulk fermions in 
addition to the massless bulk matter considered 
in the model I. The massive bulk matter contributes 
to the Higgs potential in a manner similar to the 
Boltzmann-like suppression factor, which is 
similar with finite temperature field 
theory \cite{dj}. Under the condition that the 
contribution from the massless 
bulk matter to the mass term vanishes, the mass 
term is controlled only by the term generated 
from the massive bulk fermion with the 
Boltzmann suppression factor to yield the 
exponentially small VEV. One, then, can
have the large gauge hierarchy. The flavor 
number of the massless bulk matter has 
no effect on the size of the gauge hierarchy.  
\par
%%%%%%%%%%%%
We also study the Higgs mass in each model. In the 
model I, because of the fact that the Higgs 
potential has the same form as the massless scalar field
theory studied by Coleman and Weinberg, the Higgs mass is 
inevitably smaller than massive gauge 
bosons \cite{cw}. The heavier Higgs mass tends 
to decrease the size of the gauge hierarchy. The 
large gauge hierarchy and the heavy Higgs mass 
are not compatible in the model I. On  the 
other hand, in the model II, the Higgs mass 
becomes heavier as the gauge hierarchy is 
larger. It is possible to have the consistent Higgs 
mass with the experimental lower bound for small 
flavor number. 
\par
%%%%%%%%%%%
This paper is organized as follows. In the next section we
briefly review the gauge-Higgs unification in 
the five dimensional $SU(3)$ gauge theory compactified on
the orbifold $S^1/Z_2$. We discuss the gauge hierarchy 
in the scenario of the gauge-Higgs unification and 
present two models, which can realize the large gauge 
hierarchy in section $3$. The final
section is devoted to conclusions and discussions.
%%%%%%%%%%%%%%
\section{$SU(3)$ gauge theory on $M^4\times S^1/Z_2$}
In this section, we quickly review the relevant part 
of the gauge-Higgs unification for latter 
convenience. Readers who are familiar with the 
gauge-Higgs unification can skip this section and 
directly go to the next section. We consider 
an $SU(3)$ nonsupersymmetric gauge theory 
on $M^4\times S^1/Z_2$ as the simplest example of
the gauge-Higgs unification \cite{orbi}. Here, $M^4$ is 
the four dimensional Minkowski space-time 
and $S^1/Z_2$ is an orbifold. The orbifold 
has two fixed points at $y=0, \pi R$, where 
$R$ is the radius of $S^1$. One needs to 
specify boundary conditions of fields 
for the $S^1$ direction and the fixed point.
\par
%%%%%%%
We define that
\begin{eqnarray}
A_{\hmu}(x, y + 2\pi R) &=&
U A_{\hmu}(x, y ) \, U^\dagger ,\\
\pmatrix{A_\mu \cr A_y \cr} (x, y_i - y) &=&
P_i \pmatrix{A_\mu \cr - A_y \cr} (x, y_i + y) \, 
P_i^\dagger,~~ (i = 0, 1)
\end{eqnarray}
where $U^\dagger = U^{-1}, P_i^\dagger = P_i= P_i^{-1}$ and 
$y_0=0, y_1=\pi R$ and $\hmu$ stands for $\hmu=(\mu, y)$. The 
minus sign for $A_y$ is needed to preserve the 
invariance of the Lagrangian under these transformations. 
A transformation $\pi R+y \stackrel{P_1}{\rightarrow} 
\pi R -y$ must be the same as $\pi R +y \stackrel{P_0}
{\rightarrow}-(\pi R + y)\stackrel{U}{\rightarrow} \pi R -y$, 
so we obtain $U = P_1 P_0$. Hereafter, we consider $P_i$ to be 
fundamental quantity. 
\par
%%%%%%%%%%%
For the given matrix $P_i$, the 
parity of $A_{\hmu}^a (a=1,\cdots 8)$ under 
the transformation is assigned. The fields with even parity
have zero modes, while those with odd parity have no zero modes. 
We choose 
$P_0=P_1=\e^{\pi i \sqrt{3}\lambda_8}={\rm diag.}~(-1,-1,1)$, where
$\lambda_8$ is the $8$th Gell-Mann matrix. 
Then, we observe that the zero modes are 
\begin{eqnarray}
A_{\mu}^{(0)} &=& \half \pmatrix{
A_{\mu}^{(0)3} +{A_{\mu}^{(0)8}\over\sqrt{3}} & 
A_{\mu}^{(0)1}-iA_{\mu}^{(0)2}  & 0 \cr 
A_{\mu}^{(0)1}+iA_{\mu}^{(0)2} & 
-A_{\mu}^{(0)3}+{A_{\mu}^{(0)8}\over\sqrt{3}}  & 0 \cr
0 & 0  & -{2\over\sqrt{3}}A_{\mu}^{(0)8}},\\
A_y^{(0)}&= &\half \pmatrix{
0& 0 & A_y^{(0)4}-iA_y^{(0)5} \cr 
0& 0  & A_y^{(0)6}-iA_y^{(0)7} \cr
A_y^{(0)4}+iA_y^{(0)5} & A_y^{(0)6}+iA_y^{(0)7} & 0 }.
\end{eqnarray}
Counting the zero modes for $A_{\mu}$, we see that 
the gauge symmetry is broken down 
to $SU(2) \times U(1)$ \cite{orbi}. On the other 
hand, the zero modes for $A_y$ transform as 
the $SU(2)$ doublet, so that we can 
regard it as the Higgs doublet,
\begin{equation}
\Phi\equiv \sqrt{2\pi R}~{1\over{\sqrt{2}}}~
\pmatrix{A_y^{(0)4} -i A_y^{(0)5} \cr 
A_y^{(0)6} -i A_y^{(0)7} \cr}.
\end{equation}
In fact, $\Phi$ has the $SU(2)\times U(1)$ invariant kinetic 
term \footnote{This simplest example of the gauge-Higgs
unification predicts wrong values of the Weinberg angle.
%In this simplest example of the gauge-Higgs 
%unification, the Weinberg angle is predicted to be wrong values. 
%
It is known that there are a few prescriptions to overcome 
the problem \cite{serone}.} arising from ${\rm tr}(F_{\mu y})^2$,
\begin{equation} 
\int_0^{2\pi R}dy~\biggl\{-{\rm tr}(F_{\mu y}F^{\mu y})\biggr\}=
\abs{\left(
\del_{\mu} - ig_4 A_{\mu}^{(0)a}{\tau^a\over 2}
-i {{\sqrt{3}g_4}\over 2}A_{\mu}^{(0)8}\right)\Phi}^2,
\label{shiki6}
\end{equation}
where we have rescaled the zero modes of the 
gauge fields by $\sqrt{2\pi R}$ in order 
to have the correct canonical dimension.
The VEV of the Higgs field is parametrized, by utilizing 
the $SU(2)\times U(1)$, as
\begin{equation}
\vev{A_y^{(0)}}\equiv {a\over{g_4 R}}{\lambda^6\over 2}
=A_y^{(0)6}{\lambda^6\over 2},
\end{equation}
where $a$ is a dimensionless real parameter, and $g_4$ 
is the four dimensional gauge coupling defined 
from the original five dimensional
gauge coupling by $g_4\equiv g_5/\sqrt{2\pi R}$.  
\par
%%%%%%%%
One usually evaluates the effective potential for the 
parameter $a$ in order to determine 
it \cite{hosotani}. Let us note that $a$ is closely 
related with the Wilson line phase, 
\begin{equation}
W = {\cal P} \mbox{exp} \left(ig_5 \oint_{S^1}dy \vev{A_y} \right)
= \pmatrix{
1 & 0 & 0 \cr 
0 & \cos(\pi a_0)  & i\sin(\pi a_0) \cr
0 & i\sin(\pi a_0) & \cos(\pi a_0) }\qquad (a_0~~\mbox{mod}~2),
\end{equation}
where $a_0$ is determined as the minimum 
of the effective potential. The gauge 
symmetry breaking depends on $a_0$,
\begin{equation}
SU(2)\times U(1)\rightarrow \left\{
\begin{array}{lll}
SU(2)\times U(1) & \mbox{for} & a_0=0, \\[0.1cm]
U(1)^{\prime}\times U(1)& \mbox{for} & a_0=1, \\[0.1cm]
U(1)_{{\rm em}} & \mbox{for} & \mbox{otherwise}.
\end{array}\right.
\end{equation}
\par
%%%%%%%
It has been known that in order to realize the 
desirable gauge symmetry
breaking, $SU(2)\times U(1)\rightarrow U(1)_{em}$, the 
matter content in the 
bulk is crucial. Let us consider the matter content 
studied in \cite{ghrelation} for our purpose. Following 
the standard prescription to calculate the 
effective potential 
for $a$ \cite{pomarol}\cite{hosotani}, we obtain that
\begin{eqnarray} 
V_{eff}(a) &=&
{\Gamma(5/2) \over{{\pi^{5/2}(2\pi R)^5}}} 
\sum_{n=1}^{\infty}{1 \over n^5} \nonumber\\
&\times & 
\Biggl[
\left(-3+4N_{adj}^{(+)}-dN_{adj}^{(+)s}\right)
\left(\cos[2\pi na] + 2\cos[\pi n a] \right) 
\nonumber
\\
&+&\left(4N_{adj}^{(-)}-dN_{adj}^{(-)s}\right) 
\left( \cos[2\pi n\biggl(a - \half\biggr)] 
+2 \cos[\pi n(a-1)] \right) 
\nonumber
\\
&+& \left(4N_{fd}^{(+)}-2N_{fd}^{(+)s} \right)
\cos[\pi n a] + 
\left(4N_{fd}^{(-)}- 2N_{fd}^{(-)s}\right) 
\cos[\pi n (a-1)]
\Biggr],
\label{shiki9}
\end{eqnarray}
where the factor $d$ coming from the adjoint scalar 
takes $1~(2)$ for the real
(complex) field. $N_{adj}^{(\pm)} (N_{fd}^{(\pm)})$ 
denotes the flavor
number for fermions belonging to the 
adjoint (fundamental) representation 
under $SU(3)$, where the sign $(\pm)$ stands for the
intrinsic parity defined in \cite{haba}. 
Similarly, $N_{adj}^{(\pm)s} (N_{fd}^{(\pm)s})$ means 
the flavor number for bosons in corresponding 
representations. Since $V_{eff}(a)=V_{eff}(-a)$
and $V_{eff}(a)=V_{eff}(2-a)$, physical 
region of $a$ is given by $0\leq a \leq 1$.  
\par
%%%%%%%
We have obtained the effective potential for
nonlocal quantity, the Wilson line phase $a$. 
Nonlocal terms in the effective potential are expected not to
suffer from ultraviolet effects \cite{masiero}.
%
%It is this nonlocality that the effective
%potential does not suffer from ultraviolet 
%effects \cite{masiero}. 
%
That is why the
divergence associated with the phase $a$ does 
not appear in the potential (\ref{shiki9}). The 
effective potential is calculable in the gauge-Higgs 
unification. Accordingly, the Higgs mass, which is 
obtained from the second derivative of the 
effective potential evaluated at the minimum, is 
also calculable. Hence, the gauge-Higgs unification
can provide a natural framework to address 
the gauge hierarchy problem.
%%%%%%%%%
\section{Large gauge hierarchy in the gauge-Higgs unification}
In the scenario of the gauge-Higgs unification, the mass of the 
$W$-bosons is given, from Eq. (\ref{shiki6}), by
\begin{equation} 
M_W={a_0\over {2R}}.
\end{equation}
This relation gives us the ratio between the weak scale and
the compactification scale $M_c\equiv (2\pi R)^{-1}$,
\begin{equation}
{M_W\over M_c}=\pi a_0.
\label{shiki11}
\end{equation}
Once the value of $a_0$ is determined from 
the effective potential, the compactification 
scale $M_c$ is fixed through Eq. (\ref{shiki11}). 
In the usual scenario of the gauge-Higgs 
unification, the order of $a_0$ is $O(10^{-2})$ 
for appropriate choice of the flavor set for the
massless bulk matter \cite{hty2}\cite{ghrelation}. 
Hence, the scale $M_c$ is about a few 
TeV \footnote{This is the reason that the gauge-Higgs 
unification can provide us with a solution to the 
little hierarchy problem.}. 
\par
%%%%%%%%
We would like to realize the large gauge 
hierarchy such as $M_c\sim M_{GUT},~M_{Planck}$. One 
needs very small values of $a_0$. For small 
values of $a_0$, the effective potential is 
approximated in terms of the
logarithm and polynomials with respect to $a$ by 
using the following formulae \footnote{It turns out that 
the ${\rm ln}~x$ term in 
Eq. (\ref{shiki13}) is important for the later analyses. The term
arises from the one-loop diagrams in which massless modes propagate.},
\begin{eqnarray}
\sum_{n=1}^{\infty}{1\over n^5}\cos(nx) &=&
\zeta(5)-{\zeta(3)\over 2}x^2 +
{1\over {24}}\left(-{\rm ln}~x 
+{25\over 12}\right)x^4 +O(x^6)
\label{shiki13}\\
\sum_{n=1}^{\infty}{1\over n^5}
\cos(nx-n\pi)&= &-{15\over 16}\zeta(5)
+{3\over 8}\zeta(3)x^2-
{{{\rm ln}2}\over {24}}x^4+O(x^6)
\end{eqnarray}
for $x \ll 1$ \cite{hty2}. Applying the formulae to the effective 
potential (\ref{shiki9}), one obtains that
\begin{equation}
{\bar V}_{eff}(a)= -{\zeta(3)\over 2}C^{(2)}(\pi a)^2
+{(\pi a)^4\over {24}}
\left[C^{(3)}\left(-{\rm ln}(\pi a) +{25\over 12}
\right)+C^{(4)}({\rm ln}2)
\right]+\cdots,
\label{shiki14}
\end{equation}
where $V_{eff}(a)\equiv C{\bar V}_{eff}(a)$  with 
$C\equiv {\Gamma({5\over 2})/{\pi^{5\over 2}(2\pi R)^5}}$, and the
coefficient $C^{(i)} (i=2,3,4)$ is defined by 
\begin{eqnarray}
C^{(2)}&\equiv & 24N_{adj}^{(+)}+4N_{fd}^{(+)}
+{9d\over 2}N_{adj}^{(-)s}
+{3\over 2}N_{fd}^{(-)s}\nonumber\\
&&-\left(18+6dN_{adj}^{(+)s}+2N_{fd}^{(+)s}
+18N_{adj}^{(-)}+3N_{fd}^{(-)}
\right),\label{shiki15}\\
C^{(3)}&\equiv & 72N_{adj}^{(+)}+4N_{fd}^{(+)}
-\left(54+18dN_{adj}^{(+)s}+2N_{fd}^{(+)s}\right),\\
C^{(4)}&\equiv & 48+16dN_{adj}^{(+)s}
+18dN_{adj}^{(-)s}+2N_{fd}^{(-)s}
-\left(64N_{adj}^{(+)}+4N_{fd}^{(-)}
+72N_{adj}^{(-)}\right).
\end{eqnarray}
It should be noticed that each coefficient in the 
effective potential is given by the discrete 
values, that is, the flavor number of the massless 
bulk matter. This is the very curious feature of the 
Higgs potential, which is hardly seen in the usual 
quantum field theory, and is a key point to
discuss the large gauge hierarchy in 
the gauge-Higgs unification.
%%%%%%%%%
\subsection{Large gauge hierarchy in model I}
%%%%%
%%%%
We would like to obtain a hierarchically
small VEV of $a$ as the minimum of the effective 
potential (\ref{shiki14}). It is, however, hard to realize 
such small values of $a_0$ with non-vanishing $C^{(2)}$ in the 
potential (\ref{shiki14}) \cite{hty2}.
\par
%%%%
Let us consider the case where the coefficient 
of the mass term in the Higgs potential vanishes,
\begin{equation}
C^{(2)}=0.
\label{shiki18}
\end{equation}
It should be noticed that the vanishing mass 
term (\ref{shiki18}) is not the fine
tuning of the parameter usually done in the
quantum field theory. In the present case, all 
the coefficients in the effective potential is given
by the discrete values, so that the condition 
is fulfilled just by the choice of the flavor 
set. We will discuss the matter content which 
realizes the vanishing mass term later. For 
a moment, we study the physical consequence of it. 
\par
%%%%%%%%
When Eq. (\ref{shiki18}) is satisfied, the minimum 
of the effective potential is given by 
\begin{equation}
\pi a_0 \simeq {\rm exp}\left(
{C^{(4)}\over {C^{(3)}}}~{\rm ln}2+{11\over 6}
\right) 
= {\rm exp}
\left(-{C^{(4)}\over \abs{C^{(3)}}}~{\rm ln}2
+{11\over 6}
\right).
\label{shiki19}
\end{equation}
As we will 
see later, the coefficient $C^{(3)}$ should be
negative in order for the minimum $a_0$ to 
be, at least,  a local 
minimum. Remembering Eq. (\ref{shiki11}), we have
\begin{equation} 
{M_W\over M_c}={\rm exp}
\left(-
{C^{(4)}\over \abs{C^{(3)}}}~{\rm ln}2+{11\over 6}
\right).
\label{shiki20}
\end{equation}
If we set $M_W=10^2$ (GeV) and 
$M_c=10^p$ (GeV), one obtains that
\begin{equation}
{C^{(4)}\over \abs{C^{(3)}}}={1\over{{\rm ln}2}}
\left(
{11\over 6}-(2-p){\rm ln}10
\right).
\end{equation}
The magnitude of $C^{(4)}/\abs{C^{(3)}}$ for 
various values of $p$ is listed in Table $1$.
%%%%%%%%%%%%%%%
\begin{table}[t]
$$
\begin{array}{|c||c|c|c|c|c|}\hline
 & p=11& p=12& p=13& p=16 &p=19 \\[0.3cm] \hline
{C^{(4)}\over\abs{C^{(3)}}} 
& 32.54& 35.86& 39.19& 49.15& 59.12 \\ \hline
\end{array}
$$
\caption{The magnitude of $C^{(4)}/\abs{C^{(3)}}$ for 
various values of $p$. $p=19~(16)$ corresponds 
to the Planck (GUT) scale.}
\end{table}
%%%%%%%%%new table %%%%%%
%\begin{table}[t]
%$$
%\begin{array}{|c||c|c|c|}\hline
% & p=11& p=13&p=19 \\[0.3cm] \hline
%{C^{(4)}\over\abs{C^{(3)}}} 
%& 32.54& 39.19& 59.12 \\ \hline
%\end{array}
%$$
%\caption{The magnitude of $C^{(4)}/\abs{C^{(3)}}$ for 
%various values of $p$. $p=19$ corresponds 
%to the Planck scale.}
%\end{table}
%%%%%%%%%%%%%%%%%
One requires $C^{(4)}/\abs{C^{(3)}} \gg 1$ for 
the large gauge hierarchy (large values of $p$). If 
one has the vanishing mass term in such a way 
that $C^{(4)}/\abs{C^{(3)}}$ is as large as 
listed in the table $1$, the large gauge 
hierarchy is realized in the scenario 
of the gauge-Higgs unification. 
\par
%%%%%%%%%%%
It is instructive to point out the difference 
between the present model and the famous 
Coleman-Weinberg potential of the massless 
scalar field theory \cite{cw}. The effective 
potential with $C^{(2)}=0$ (called model I 
hereafter) is exactly the same form as the one 
in the paper by Coleman and Weinberg \cite{cw}. The 
potential is controlled by the logarithmic and 
quartic terms. There is, however, a big difference 
among them. In the Coleman-Weinberg's case, the 
quartic coupling exists at the tree-level and 
only the logarithmic term is generated at the 
one-loop level. The logarithmic term becomes 
dominant contribution against the quartic term 
at the nontrivial vacuum configuration, so that 
the vacuum configuration is outside of the validity of 
perturbation theory. On the other hand, for 
the present case, both of the logarithmic and 
quartic terms are generated at the one-loop
level, even if the quartic and the logarithmic 
terms are the same order to each other at 
the nontrivial vacuum 
configuration (\ref{shiki19}), the perturbative 
reliability for the vacuum configuration is 
not spoiled. 
%
%Rather, what one has to take care 
%is 
Rather, what one has to be concerned with is 
the stability against the two (higher) loop 
contributions \footnote{One, of course, cares about 
the vanishing mass term (\ref{shiki18}) at the 
two (higher)-loop level as well. We will come 
back to this point as well later .}. We will 
discuss this point later.      
\par
%%%%%%%%%%%% 
Since we understand that it is possible to have 
the large gauge hierarchy in the scenario 
of the gauge-Higgs unification, let us next present
explicit examples of the flavor set to realize 
it. To this end, we investigate the 
condition (\ref{shiki18}), which is a key ingredient 
for the large gauge hierarchy.
\par
%%%%%%%%%%
We recast Eq. (\ref{shiki15}) as
\begin{eqnarray} 
C^{(2)}&=& 6\left(4N_{adj}^{(+)}-3-dN_{adj}^{(+)s}\right)
+2\left(2N_{fd}^{(+)}-N_{fd}^{(+)s}\right)\nonumber\\
&&+{9\over 2}\left(dN_{adj}^{(-)s}-4N_{adj}^{(-)}\right)
+{3\over 2}\left(N_{fd}^{(-)s}-2N_{fd}^{(-)}\right).
\label{shiki22}
\end{eqnarray}
In order to fulfill $C^{(2)}=0$, the second 
parenthesis $(2N_{fd}^{(+)}-N_{fd}^{(+)s})$ in 
Eq. (\ref{shiki22}) must be an integral 
multiple of $3$. Accordingly, 
\begin{equation}
C^{(3)}=18\left(4N_{adj}^{(+)}-3-dN_{adj}^{(+)s}\right)
+2\left(2N_{fd}^{(+)}-N_{fd}^{(+)s}\right) 
\end{equation}
is an integral multiple of $6$. Then, we write
\begin{equation}
C^{(3)}= -6k \left(= -\abs{C^{(3)}}\right)
\qquad (k={\rm positive~~integer}).
\end{equation}
Let us introduce another integer $m$ defined by
\begin{equation} 
4N_{adj}^{(+)}-3-dN_{adj}^{(+)s}\equiv -m.
\label{shiki25}
\end{equation}
From the coefficient $C^{(3)}$, we have 
\begin{equation}
2N_{fd}^{(+)}-N_{fd}^{(+)s}=-3k+9m.
\label{shiki26}
\end{equation}
%%%%%%%%
Imposing the condition $C^{(2)}=0$ gives us a 
relation given by
\begin{equation}
N_{fd}^{(-)s}-2N_{fd}^{(-)}=
-3\left(dN_{adj}^{(-)s}-4N_{adj}^{(-)}\right)
+4k-8m,
\label{shiki27}
\end{equation}
where we have used Eqs. (\ref{shiki25}) 
and (\ref{shiki26}). Equipped with these 
equations, we obtain that
\begin{equation}
C^{(4)}=12\left(dN_{adj}^{(-)s}-4N_{adj}^{(-)}\right)+8k,
\label{shiki28}
\end{equation}
which is independent of the 
integer $m$. Hence, we finally have
\begin{equation}
{C^{(4)}\over\abs{ C^{(3)}}}={2\over k}
\left(dN_{adj}^{(-)s}-4N_{adj}^{(-)}\right)+{4\over 3}.
\label{shiki29}
\end{equation}
This result tells us an important point 
that in order to make $C^{(4)}/\abs{C^{(3)}}$ 
larger, smaller values of $\abs{C^{(3)}}$ 
and $N_{adj}^{(-)}$ are favored. In fact, $k=1$ is 
the most desirable choice for the large gauge hierarchy.
\par
%%%%%%%%%%%
Let us now study the matter content in the 
model I. For a given $p$, the values 
for $C^{(4)}/\abs{C^{(3)}}$ is determined. Then, we 
obtain a flavor set $(N_{adj}^{(-)}, dN_{adj}^{(-)s})$ 
for a fixed values of $k$ through Eq. (\ref{shiki29}).
One also fixes the integer $m$. Then, a 
flavor set $(N_{fd}^{(-)}, N_{fd}^{(-)s})$ is
determined by Eq.(\ref{shiki27}) and 
also $(N_{fd}^{(+)}, N_{fd}^{(+)s})$ by 
Eq. (\ref{shiki26}). Finally, one obtains 
$(N_{adj}^{(+)}, dN_{adj}^{(+)s})$ from 
Eq. (\ref{shiki25}). We notice that the flavor 
number of the massless bulk matter 
with $(+)$ parity depends only on the values 
of $k$ and $m$, but not depends on the values 
of $p$, as seen from Eqs. (\ref{shiki25}) 
and (\ref{shiki26}). The flavor number of 
the massless bulk matter with 
the $(-)$ parity depends on the values of $p$, which 
sets the hierarchy between $M_W$ and $M_c$.
\par
%%%%%%%%%%%
Let us present a few examples of the flavor set 
in the model I. We choose $(k, m)=(1,0)$ as 
an example. Then, we find that 
\begin{eqnarray}
4N_{adj}^{(+)}-3-dN_{adj}^{(+)s}=0 &\rightarrow &(N_{adj}^{(+)},
dN_{adj}^{(+)s})=(1, 1), (2, 5), (3, 9),\cdots,\\
2N_{fd}^{(+)}-N_{fd}^{(+)s}=-3 &\rightarrow & 
(N_{fd}^{(+)}, N_{fd}^{(+)s})=(0, 3), (1, 5), (2, 7), \cdots.
\end{eqnarray}
\par\noindent
%%%%%%%%
For $(k, m, p)=(1, 0, 19)$, 
\begin{eqnarray}
dN_{adj}^{(-)s}-4N_{adj}^{(-)}\simeq 29 &\rightarrow&
(N_{adj}^{(-)}, dN_{adj}^{(-)s})=(0, 29), (1, 33), (2, 37),\cdots \\
N_{fd}^{(-)s}-2N_{fd}^{(-)}=-83 &\rightarrow & 
(N_{fd}^{(-)}, N_{fd}^{(-)s})=(42, 1), (43, 3), (44, 5), \cdots.
\end{eqnarray}
\par\noindent
%%%%%%%%%%%%%
%For $(k, m, p)=(1, 0, 16)$, 
%\begin{eqnarray}
%dN_{adj}^{(-)s}-4N_{adj}^{(-)}\simeq 24 &\rightarrow&
%(N_{adj}^{(-)}, dN_{adj}^{(-)s})=(0, 24), (1, 28), (2, 32),\cdots \\
%N_{fd}^{(-)s}-2N_{fd}^{(-)}=-68 &\rightarrow & 
%(N_{fd}^{(-)}, N_{fd}^{(-)s})=(34, 0), (35, 2), (36, 4), \cdots.
%\end{eqnarray}
%\par\noindent
%%%%%%%%%%
%For $(k, m, p)=(1, 0, 13)$, 
%\begin{eqnarray}
%dN_{adj}^{(-)s}-4N_{adj}^{(-)}\simeq 19 &\rightarrow&
%(N_{adj}^{(-)}, dN_{adj}^{(-)s})=(0, 19), (1, 23), (2, 27),\cdots \\
%N_{fd}^{(-)s}-2N_{fd}^{(-)}=-53 &\rightarrow & 
%(N_{fd}^{(-)}, N_{fd}^{(-)s})=(27, 1), (28, 3), (29, 5), \cdots.
%\end{eqnarray}
%\par\noindent
%%%%%%%%%%
%For $(k, m, p)=(1, 0, 12)$, 
%\begin{eqnarray}
%dN_{adj}^{(-)s}-4N_{adj}^{(-)}\simeq 17 &\rightarrow&
%(N_{adj}^{(-)}, dN_{adj}^{(-)s})=(0, 17), (1, 21), (2, 25),\cdots \\
%N_{fd}^{(-)s}-2N_{fd}^{(-)}=-48 &\rightarrow & 
%(N_{fd}^{(-)}, N_{fd}^{(-)s})=(24, 0), (25, 2), (26, 4), \cdots.
%\end{eqnarray}
%\par\noindent
%%%%%%%%%%
For $(k, m, p)=(1, 0, 11)$, 
\begin{eqnarray}
dN_{adj}^{(-)s}-4N_{adj}^{(-)}\simeq 16 &\rightarrow&
(N_{adj}^{(-)}, dN_{adj}^{(-)s})=(0, 16), (1, 20), (2, 24),\cdots \\
N_{fd}^{(-)s}-2N_{fd}^{(-)}=-43 &\rightarrow & 
(N_{fd}^{(-)}, N_{fd}^{(-)s})=(22, 1), (23, 3), (24, 5), \cdots.
\end{eqnarray}
%%%%%%%%%%
We observe that the flavor 
numbers $dN_{adj}^{(-)s}, N_{fd}^{(-)}$ are of 
order of $O(10)$. One should worry about 
%
%One has to take care about 
%
the reliability of perturbation theory for such the 
large number of flavor. This is because an expansion 
parameter in the present case may be given by
$(g_4^2/4\pi^2)N_{flavor}$, and it must 
be $(g_4^2/4\pi^2)N_{flavor} \ll 1$ for 
reliable perturbative expansion. 
\par
%%%%%%%%%%%
Now, let us discuss the Higgs mass in the
model I. The Higgs mass squared is obtained 
by the second derivative of the effective 
potential evaluated at the minimum of the 
potential. We have 
\begin{equation}
m_H^2=(g_5R)^2~C~
{{\del^2 {\bar V}_{eff}}\over {\del a^2}}\bigg|_{a=a_0}
={{3g_4^2}\over {16\pi^2}}~M_W^2~\left(-{C^{(3)}\over 6}\right)
\left(={{3g_4^2}\over {16\pi^2}}~M_W^2~k < M_W^2\right),
\label{shiki42}
\end{equation}
where we have used Eq. (\ref{shiki19}). This 
shows that, as we have stated before, the 
coefficient $C^{(3)}$ must be negative in 
order for the minimum $a_0$ to be, at least, a 
local minimum of the effective potential. The 
larger values of $C^{(3)}$ makes the Higgs 
mass heavier, while, as we have discussed, the 
large gauge hierarchy favors smaller 
values of $C^{(3)}$. The choice $k=1$ 
is the most desirable one for the 
large gauge hierarchy, so that the Higgs 
mass is lighter than $M_W$, which is the same 
result in the original Coleman-Weinberg's 
paper \cite{cw}. Therefore, one concludes 
that the large gauge hierarchy and 
the sufficiently heavy Higgs mass are not 
compatible in the model I.   
%%%%%%%%%%%%%%%%%%
\subsection{Large gauge hierarchy in model II}
In this subsection, we study another model called 
model II. In the model II, we introduce 
massive bulk fermions \cite{takenaga}--\cite{marutake}, that 
is, fermions with bulk mass term in addition to 
the massless bulk matter in the model I. As is well 
known, the bulk mass term for fermion in five 
dimensions is odd under the parity 
transformation, $y\rightarrow 
-y (\pi R-y \rightarrow \pi R+y)$. 
One needs parity-even mass term for
the consistency of the $Z_2$ orbifolding. 
We resort to one of the prescriptions known 
to realize such the mass term. Here, we introduce 
a pair of fermion fields, $\psi^{(+)}, \psi^{(-)}$. 
$\psi^{(+)}$ and $\psi^{(-)}$ have the different 
$Z_2$ parity, so that the 
mass term $M{\bar\psi}^{(+)}\psi^{(-)}$ has even 
parity under the $Z_2$. A detailed discussion is 
given in \cite{marutake}. 
\par
%%%%%%%%%%
Then, the contribution to the
effective potential from the massive fermions 
is given by
\begin{equation}
V_{eff}^{massive}=-2^{[5/2]}(1+1)
N_{pair}{1\over 2\pi R}{1\over 2}
\sum_{n=-\infty}^{\infty}\int {{d^4p_E}\over (2\pi)^4}~
{\rm ln}\left[
p_E^2 +\left({{n+{a\over 2}}\over R}\right)^2 +M^2
\right],
\end{equation}
where we have assumed the fermions belong to the 
fundamental representation under the gauge 
group $SU(3)$ and $N_{pair}$ stands for the number 
of the pair $(\psi^{(+)}, \psi^{(-)})$. According 
to the usual prescription \cite{pomarol}, we have 
\begin{equation}
V_{eff}^{massive}={3\over{4\pi^2 (2\pi R)^5}}~
4\times 2N_{pair}\sum_{n=1}^{\infty}{1\over n^5}
\left(
1+nz+{n^2z^2\over 3}
\right)\e^{-nz}\cos(\pi n a),
\label{shiki44}
\end{equation}
where we have defined a dimensionless 
parameter $z\equiv 2\pi RM=M/M_c$.
\par
%%%%%%
We are interested in the very small values 
of $a$, so that we expand the cosine function 
to obtain that
\begin{equation}
V_{eff}^{massive}=
{3\over{4\pi^3 (2\pi R)^5}}4\times 2N_{pair}
\left[
B^{(0)}-B^{(2)}{\pi^2\over 2}a^2 
+ B^{(4)}{\pi^4\over 4!}a^4+\cdots
\right],
\end{equation}
where
\begin{eqnarray}
B^{(0)}&=&\sum_{n=1}^{\infty}{1\over n^5}
\left(1+nz+{n^2z^2\over 3}\right)\e^{-nz},\\
B^{(2)}&=&\sum_{n=1}^{\infty}{1\over n^3}
\left(1+nz+{n^2z^2\over 3}\right)\e^{-nz},\\
B^{(4)}&=&\sum_{n=1}^{\infty}{1\over n}
\left(1+nz+{n^2z^2\over 3}\right)\e^{-nz}.
\end{eqnarray}
We note that the coefficient $B^{(i)} (i=0, 2, 4)$ is 
suppressed by the Boltzmann-like 
factor $\e^{-nz}$, reflecting the fact that the 
effective potential (\ref{shiki44}) shares 
similarity with that in finite temperature 
field theory \cite{dj}.
\par
%%%%%%%%%
The total effective potential to the Wilson 
line phase $a$ (Higgs field) is given by 
Eqs. (\ref{shiki14}) and (\ref{shiki44}),
\begin{eqnarray}
{\bar V}_{eff}^{total}&\simeq & 
-\half\left[\zeta(3) C^{(2)}
+8N_{pair}B^{(2)}\right](\pi a)^2
\nonumber\\
&&+{(\pi a)^4\over {24}}
\left[C^{(3)}\left(-{\rm ln}(\pi a) +{25\over 12}
\right)+C^{(4)}({\rm ln}2)
\right],
\end{eqnarray}
where we have ignored the contribution to the 
quartic term from the massive bulk fermion 
because it is highly suppressed. We
require, again, that  the contribution to the
mass term from the massless bulk matter 
vanishes, that is, $C^{(2)}=0$. The mass 
term is, then, controlled only by the term 
coming from the massive bulk fermion. The 
magnitude of the VEV is governed by the
mass term in the Higgs potential, so that the 
VEV for the present case is exponentially 
suppressed for appropriate large values 
of $z (> 1)$. The essential behavior of the VEV
is governed by the factor $B^{(2)}$, {\it i.e.}
\begin{equation}   
\pi a_0 \simeq \gamma B^{(2)}
\end{equation}
with some numerical constant $\gamma$ of order $1$.
%%%%%%%%%
Since we assume $z=M/M_c >1$, the size of the bulk
mass parameter $M$ may be determined by physics above
the compactification scale $M_c$. If we take $M$ to 
be the order of the cutoff scale, it is natural to
assume that $M_c$ is, at most, at around $10^{17}$ GeV in 
the model II. 
%%%%%%%% 
%Since $z > 1$, the size of the bulk mass 
%parameter $M$ may be determined by physics above 
%the compactification scale $M_c$. It is natural to
%consider $M_c\leq M_{Planck}$, so that we assume 
%$M_c$ is, at most, at around $10^{17}$ GeV in 
%the model II.   
%%%%%%%
\par
%%%%%
If we write $\pi a_0 = \e^{-Y}$, then, one 
finds, remembering Eq. (\ref{shiki11}), that
\begin{equation}
-Y={\rm ln}(\pi a_0)\left(={\rm ln}
\left({M_W\over M_c}\right)\right)=(2-p)
{\rm ln}10 \simeq 
\left\{
\begin{array}{ccc}
%-39.144 & \mbox{for} & p=19, \\
%-36.841 & \mbox{for} & p=18, \\
-34.539 & \mbox{for} & p=17, \\
-32.236 & \mbox{for} & p=16, \\
-25.328 & \mbox{for} & p=13, \\
-23.026 & \mbox{for} & p=12, \\
-20.723 & \mbox{for} & p=11.
\end{array}\right.
\end{equation}
The gauge hierarchy is controlled by the magnitude 
of $Y$, in other words, the bulk mass 
parameter $z$, and the large gauge hierarchy is 
achieved by $\abs{z}\simeq 30 \sim 40$. The 
large gauge hierarchy is realized by the 
presence of the massive bulk fermion under the 
condition (\ref{shiki18}). We notice that the 
flavor number of the massless bulk matter 
is not essential for the large gauge hierarchy 
in the model II. 
\par
%%%%%%%%%%
Now, let us next discuss the Higgs mass 
in the model II. Again, the Higgs mass 
is obtained by the second 
derivative of the total effective potential 
at the vacuum configuration $a_0$,
\begin{eqnarray}
m_H^2&=&(g_5R)^2~C~
{{\del^2 {\bar V}_{total}}\over {\del a^2}}
\bigg|_{a=a_0}\nonumber\\
&=&{g_4^2\over{16\pi^2}}M_W^2
\left[
-C^{(3)}{\rm ln}(\pi a_0)+{4\over 3}C^{(3)}
+C^{(4)}{\rm ln}2
\right]\nonumber\\
&=&{g_4^2\over{16\pi^2}}M_W^2~F,
\end{eqnarray}
where we have defined 
\begin{equation}
F\equiv -C^{(3)}{\rm ln}(\pi a_0)+{4\over 3}C^{(3)}
+C^{(4)}{\rm ln}2.
\label{shiki53}
\end{equation}
At first glance, one may think that the Higgs
mass is lighter than $M_W$ as in the case of 
the model I. This is, however, not the case 
in the model II. 
%%%%%%%%%%%
%We have the flavor number 
%degrees of freedom to enhance the Higgs mass without spoiling the 
%large gauge hierarchy and the validity of perturbation theory. 
%%%%%%%%%%%%
The Higgs mass depends on the logarithmic 
factor. The larger the gauge hierarchy 
is, the heavier the Higgs mass is. An important 
point is that the coefficient $C^{(3)}$ is 
not related with the realization of the 
large gauge hierarchy, so that it is not 
constrained by the requirement of the 
large gauge hierarchy at all. 
%%%
%In the model I, the small $C^{(3)}$ is favored for the 
%large hierarchy (\ref{shiki20}), while the 
%heavy Higgs mass requires the large $C^{(3)}$ (\ref{shiki42}). 
%%%
\par
%%%%%%%%%
The vacuum configuration must be, at least, a local 
minimum. We require that $F\geq 0$. Defining 
\begin{equation}
l\equiv dN_{adj}^{(-)s}-4N_{adj}^{(-)}
\label{shiki54}
\end{equation}
and 
recalling $\pi a_0 =\e^{-Y}$, we 
have, from Eq. (\ref{shiki53}), 
\begin{equation}
F=k\biggl(-6Y +8({\rm ln}2-1)\biggr)
+l\times 12{\rm ln}2,
\label{shiki55}
\end{equation}
where we have 
used $C^{(3)}=-6k~~(k\in {\bf Z})$ and 
Eq. (\ref{shiki28}). In the model II the 
coefficient $C^{(3)}$ is not 
necessarily negative. We separately discuss the 
size of the Higgs mass, depending in the sign of $C^{(3)}$.
%%%%%%%%
% Let us define the
%integer $l$ for latter convenience, 
%Then, we have, writing $\pi a_0 =\e^{-Y}$, 
%%%%%%%%%%%
\par
%%%%%%%%
For $C^{(3)} < 0~~(k>0)$, in order for $m_H^2$ to be 
positive, one needs that
\begin{equation}
l > k\times {{6Y + {8(1-{\rm ln}2)}}\over {12{\rm ln}2}}
\simeq k\times \left\{
\begin{array}{ccc}
%28.53 & \mbox{for} & p=19,\\
%26.87 & \mbox{for} & p=18,\\
25.21 & \mbox{for} & p=17,\\
23.55 & \mbox{for} & p=16,\\
18.57 & \mbox{for} & p=13,\\
16.90 & \mbox{for} & p=12,\\
15.24 & \mbox{for} & p=11.
\end{array}
\right.
\end{equation}
This shows that we need $O(10)$ numbers of the 
flavor for $dN_{adj}^{(-)s}$. The reliability of 
perturbation theory may be lost for such the large
number of flavor. Hence, we exclude the case of $C^{(3)}<0$ 
in the model II. Hereafter, we restrict ourselves 
to $C^{(3)}>0$, that is, $k<0$.
\par
%%%%%%%%%%
In addition to the sign of $C^{(3)}$, the integer $l$ 
can take both sign. Let us first 
consider $l < 0$. Writing $l~(k)=-\abs{l}~(-\abs{k})$, the 
requirement of $m_H^2 >0$ yields 
\begin{equation} 
\abs{k} > {12{\rm ln}2 \over 6Y 
+ 8(1-{\rm ln}2)}\times \abs{l}
\simeq 
\abs{l}\times \left\{
\begin{array}{ccc}
%0.035 & \mbox{for} & p=19,\\
%0.037 &\mbox{for} & p=18,\\
0.040  &\mbox{for} & p=17, \\
0.042 & \mbox{for} & p=16,\\
0.054 & \mbox{for} & p=13,\\
0.059 & \mbox{for} & p=12,\\
0.066 & \mbox{for} & p=11.
\end{array}
\right.
\end{equation}
Since the minimum values of $\abs{k}$ is given 
by $\abs{k}=1$, {\it i.e.}, $k=-1$, one obtains that
\begin{equation}
0 < \abs{l}~\le~
\left\{
\begin{array}{ccc}
%28 & \mbox{for} & p=19,\\
%26 &\mbox{for} & p=18,\\
25 &\mbox{for} & p=17,\\
23 & \mbox{for} & p=16,\\
18 & \mbox{for} & p=13,\\
16 & \mbox{for} & p=12,\\
15 & \mbox{for} & p=11.
\end{array}\right.
\end{equation}
The upper bound of $\abs{l}$ is larger if we 
choose larger values of $\abs{k}$. In 
order to avoid the large flavor number, let 
us choose $l=-1$ as an example. And we 
impose the constraint on the Higgs mass 
from the experimental 
lower bound, $m_H~\rough{>}~114$ GeV. Then, we
find the possible values of $k$,
\begin{equation}
\abs{k}~ \rough{>}~
{1\over{6Y+8(1-{\rm ln}2)}}
\left({16\pi^2\over g_4^2 M_W^2}~
(114~\mbox{GeV})^2+12\abs{l}{\rm ln}2\right)
\simeq 
\left\{
\begin{array}{ccc}
%3.23 & \mbox{for} & p=19,\\
%3.43 & \mbox{for} & p=18,\\
3.64 & \mbox{for} & p=17,\\
3.90 & \mbox{for} & p=16,\\
4.95 & \mbox{for} & p=13,\\
5.43 & \mbox{for} & p=12,\\
6.02 & \mbox{for} & p=11,
\end{array}\right.
\label{shiki}
\end{equation}
where we have used $g_4^2\simeq 0.42$. If we 
choose the larger values 
of $\abs{l}$, the possible values of $k$
becomes larger. We observe that the large $p$ suppresses 
the values of $k$, which means that the lower flavor number 
is enough for the Higgs mass to satisfy the experimental 
lower bound.
\par
%%%%%%%
Let us present the set of the flavor number 
for $(k, l)=(-4, -1)$. There is another free 
integer $m$, which is defined by Eq. (\ref{shiki25}). We 
choose $m=0$ as a demonstration. Then, from Eqs. (\ref{shiki25}), 
(\ref{shiki26}), (\ref{shiki27}) and (\ref{shiki54})
we have 
\begin{eqnarray}
(N_{adj}^{(-)}, dN_{adj}^{(-)s})&=&(1, 3),~(2, 7),\cdots,\label{shiki60}\\
(N_{fd}^{(-)}, N_{fd}^{(-)s})&=&(7, 1),~(8, 3),\cdots,\label{shiki61}\\
(N_{fd}^{(+)}, N_{fd}^{(+)s})&=&(6, 0),~(7, 2),\cdots,\label{shiki62}\\
(N_{adj}^{(+)}, dN_{adj}^{(+)s})&=&(1, 1),~(2, 5),\cdots.\label{shiki63}
\end{eqnarray}
The Higgs mass in GeV unit is calculated as 
\begin{eqnarray}
m_H^2 &=& {g_4^2\over 16\pi^2}~M_W^2~
\Biggl(
\abs{k}\biggl(6Y+8(1-{\rm ln}2)\biggr)-\abs{l}\times 12{\rm ln}2
\Biggr)\bigg|_{k=-4,~l=-1}\nonumber\\
\rightarrow m_H &\simeq &
\left\{
\begin{array}{ccc}
%127.0 & \mbox{for} & p=19,\\
%123.2 &\mbox{for} & p=18,\\
119.5 &\mbox{for} & p=17,\\
115.5 & \mbox{for} & p=16,\\
102.4 & \mbox{for} & p=13,\\
97.6 & \mbox{for} & p=12,\\
92.6 & \mbox{for} & p=11,
\end{array}\right.
\label{shiki64}
\end{eqnarray}
where we have used $g_4^2\simeq 0.42$ \footnote{In the 
usual scenario of the gauge-Higgs unification, one 
requires $g_4\sim O(1)$ in order to have the heavy 
enough Higgs mass \cite{ghrelation}\cite{hty2}. Our scenario can
give heavy Higgs masses compatible with experiments, even for the
weak coupling.
%
%The large gauge hierarchy enhances the Higgs mass 
%sizably even for the weak coupling.
%
}. We observe 
that for the fixed integers $(k, l)$, the large 
gauge hierarchy, that is, large ${\rm ln}(\pi a_0)=-Y$ 
enhances the size of the Higgs mass.
\par
%%%%%%%%%%%%%%%
Let us next consider the non-negative $l$ with $k <0$. In 
this case it is obvious from Eq. (\ref{shiki55}) that 
the Higgs mass is positive definite. We first consider the 
$l=0$ case. The requirement of $m_H~\rough{>}~114$ GeV
gives us the allowed values of $k=-\abs{k}$,
\begin{equation}
\abs{k}~\rough{>}~
{16\pi^2 \over g_4^2 M_W^2}{(114~\mbox{GeV})^2\over 
6Y+8(1-{\rm ln}2)}\nonumber\\
=\left\{
\begin{array}{ccc}
%3.19& \mbox{for} & p=19,\\
%3.39& \mbox{for} & p=18,\\
3.61& \mbox{for} & p=17,\\
3.87& \mbox{for} & p=16,\\
4.91& \mbox{for} & p=13,\\
5.39& \mbox{for} & p=12,\\
5.98& \mbox{for} & p=11.
\end{array}\right.
\end{equation}
The set of flavor number for $(k, l)=(-4, 0)$ with $m=0$ 
is given by
\begin{eqnarray}
(N_{adj}^{(-)}, dN_{adj}^{(-)s})&=&(1, 4),~(2, 8),\cdots,\label{shiki66}\\
(N_{fd}^{(-)}, N_{fd}^{(-)s})&=&(8, 0),~(9, 2),\cdots,\label{shiki67}\\
(N_{fd}^{(+)}, N_{fd}^{(+)s})&=&(6, 0),~(7, 2),\cdots,\label{shiki68}\\
(N_{adj}^{(+)}, dN_{adj}^{(+)s})&=&(1, 1),~(2, 5),\cdots.\label{shiki69}
\end{eqnarray}
%%%%%%%%%%
%Let us note that the flavor number for the bulk field with $(+)$ parity
%is determiend once one fixes the integers $k, m$. 
%%%%%%%%%%%
The Higgs mass in GeV unit is obtained as
\begin{eqnarray}
m_H^2 &=& {g_4^2\over 16\pi^2}~M_W^2~
\Biggl(
\abs{k}\biggl(6Y+8(1-{\rm ln}2)\biggr)+l\times 12{\rm ln}2
\Biggr)\bigg|_{k=-4,~l=0}\nonumber\\
\rightarrow m_H &\simeq & 
\left\{
\begin{array}{ccc}
%127.6 & \mbox{for} & p=19,\\
%123.8 & \mbox{for} & p=18,\\
119.9 & \mbox{for} & p=17,\\
115.9 & \mbox{for} & p=16,\\
102.9 & \mbox{for} & p=13,\\
98.2 & \mbox{for} & p=12,\\
93.2 & \mbox{for} & p=11.
\end{array}\right.
\label{shiki70}
\end{eqnarray}
We observe again that for the fixed integers $(k, l)$ the 
large gauge hierarchy enhances the size of the Higgs mass 
thanks to the large ${\rm ln}(\pi a)$. We also confirm that 
the dominant contribution to the Higgs mass is given by
${\rm ln}(\pi a)$ if we compare Eq. (\ref{shiki64}) 
with Eq. (\ref{shiki70}).
\par
%%%%%%%%%%%
%The Higgs mass is mainly controlled by the logarithmic 
%term, and in fact as we have seen, the large gauge 
%hierarchy yielding the large logarithmic term enhances 
%the size of the Higgs mass. 
%%%%%%%%%%
If $k=0$, there is no logarithmic term, and the Higgs mass 
is inevitably light. Instead of showing the Higgs mass 
in the $k=0$ case, we rather show that many flavor numbers 
are necessary to enhances the size of the Higgs mass 
though such a large flavor number violates the validity of
perturbation theory.
%
%such the large number of flavor is out of the 
%$validity of perturbation theory. 
%
For $k=0$, the Higgs mass is reduced to
\begin{equation}
m_H^2\bigg|_{k=0}={g_4^2\over 16\pi^2}M_W^2 
\times l\times 12{\rm ln}2.
\end{equation}
The allowed values of $l$ consistent 
with $m_H~\rough{>}~114$ GeV is $l~\rough{>}~92$. One 
needs many flavor numbers which should be avoided.
\par
%%%%%%%%%%
We have assumed $m=0$ in the above two 
examples, $(k, l)=(-4, -1), (-4, 0)$. If we take 
another values for $m$, we can further reduce 
the number of flavor, which is desirable
from the point of view of perturbation theory. Let 
us show the result when we take $m=-1$. The 
equations (\ref{shiki60})--(\ref{shiki63}) change to
\begin{eqnarray}
(N_{adj}^{(-)}, dN_{adj}^{(-)s})&=&(1, 3),~(2, 7),\cdots,\\
(N_{fd}^{(-)}, N_{fd}^{(-)s})&=&(3, 1),~(4, 3),\cdots,\\
(N_{fd}^{(+)}, N_{fd}^{(+)s})&=&(2, 1),~(3, 3),\cdots,\\
(N_{adj}^{(+)}, dN_{adj}^{(+)s})&=&(1, 0),~(2, 4),\cdots.
\end{eqnarray}
We observe that the flavor 
numbers $N_{fd}^{(\pm)}, N_{fd}^{(\pm)s}$ are
reduced. Likewise, the 
equations (\ref{shiki66})--(\ref{shiki69}) changes to
\begin{eqnarray}
(N_{adj}^{(-)}, dN_{adj}^{(-)s})&=&(1, 4),~(2, 8),\cdots,\\
(N_{fd}^{(-)}, N_{fd}^{(-)s})&=&(4, 0),~(5, 2),\cdots,\\
(N_{fd}^{(+)}, N_{fd}^{(+)s})&=&(2, 1),~(3, 3),\cdots,\\
(N_{adj}^{(+)}, dN_{adj}^{(+)s})&=&(1, 0),~(2, 4),\cdots.
\end{eqnarray}
Again, the flavor 
numbers $N_{fd}^{(\pm)}, N_{fd}^{(\pm)s}$ are reduced.
The Higgs mass does not depend on the 
integer $m$, so that the results Eqs. (\ref{shiki64}) 
and (\ref{shiki70}) do not change.
\par
%%%%%%%%
In the model II, the large gauge hierarchy and 
the heavy Higgs mass are compatible. The massive 
bulk fermion plays the role to generate the large 
gauge hierarchy. Once the large gauge hierarchy is 
achieved, the consistent Higgs mass can be obtained 
for the fixed set of the reasonable flavor number. The 
larger the gauge hierarchy is, the heavier the Higgs 
mass tends to be.
%%%%%%%%%%
\section{Conclusions and discussions}
We have considered the five dimensional 
nonsupersymmetric $SU(3)$ model compactified 
on $M^4\times S^1/Z_2$, which is the simplest 
model to realize the scenario of the gauge-Higgs 
unification. We have discussed whether the large 
gauge hierarchy is achieved in the scenario or not. 
The Higgs potential is generated at the one-loop 
level and is obtained in a finite form, reflecting 
the nonlocal nature that the Higgs field is the 
Wilson line phase in the gauge-Higgs 
unification. The Higgs potential is calculable 
and accordingly, so is the Higgs mass.
\par
%%%%%%%%%%%
The Higgs potential is expanded in terms of the 
logarithm and the polynomials for the small VEV 
of the Higgs field. Then, the coefficient is 
written in terms of the flavor number of the 
massless bulk matter introduced 
into the theory. This means that the coefficient 
is the discrete value, unlike the usual quantum field 
theory, in which the coefficient is a scale-dependent 
parameter. It is possible to have 
the vanishing mass term in the Higgs potential by
choosing the appropriate set of the flavor 
number. This is not the fine tuning of the 
parameters. We have found two 
models (model I, II), in which the large
gauge hierarchy is realized.  
\par
%%%%%%%%%%
In the model I, the Higgs potential consists 
of the logarithm and quartic terms, both of 
which are generated at the one-loop level. Perturbation 
theory is not spoiled even in the nontrivial vacuum
configuration, for which the size of both terms is 
the same order. This is a different point from the 
Coleman-Weinberg \cite{cw}, in which the nontrivial 
vacuum is outside of the validity of perturbation 
theory. We have found that in order
to realize the large gauge hierarchy we 
need the large (small) $C^{(4)} (C^{(3)})$. 
This, in turn, requires the $O(10)$ numbers of 
the flavor. From a point of view of perturbation 
theory, such the large flavor number is not favored. 
\par
%%%%%%%
The Higgs mass in the model I is inevitably light, lighter 
than $M_W$, which is the same result as that in 
the Coleman-Weinberg. The small $C^{(3)}$ tends to
realize the large gauge hierarchy, while the heavy 
Higgs mass needs the large $C^{(3)}$. Therefore, in 
the model I, the large gauge hierarchy is 
realized, but the Higgs mass is too light.
\par
%%%%%%%%%%%%%%%%%
In the model II, we have considered 
the massive bulk fermion in addition to the 
massless bulk matter. If we have assumed that the 
contribution to the mass term in the Higgs 
potential from the massless bulk matter 
vanishes, the dominant contribution to the 
mass term comes form the massive bulk
fermion alone whose coefficient is given by 
the Boltzmann-like suppression factor. As 
a result, the VEV of the Higgs field is 
exponentially small and the large gauge 
hierarchy is realized for the appropriate 
values of $z=M/M_c$. The flavor number of the 
massless bulk matter does not concern the gauge hierarchy. 
\par
%%%%%%%%
It is interesting to note that 
the Higgs mass in the model II becomes heavier 
%%%%%
%as the size of the gauge hierarchy is larger. It is interesting to note
%that the Higgs mass becomes heavier 
%%%%%%%%%
as the compactified scale $R$ is
smaller. We have shown that the Higgs mass can be consistent 
with the experimental lower bound without 
introducing many flavor numbers. 
%%%%%%%%%%%
\par
%%%%%%%%%%%%
We have introduced the scalar fields in the 
bulk. In general, scalar fields receive large 
radiative corrections. One might ask whether the large  
gauge hierarchy can be realized without scalar fields in the bulk.
%
%One may wonder whether the
%large gauge hierarchy is realized or not 
%only by fermions in the bulk. 
%
Unfortunately, one can neither realize the large 
gauge hierarchy nor have the stable nontrivial vacuum in the 
model I. The relevant quantity for the large 
gauge hierarchy is reduced to 
\begin{equation}  
{C^{(4)}\over \abs{C^{(3)}}}\rightarrow 
{2\over k}\left(-4N_{adj}^{(-)}\right) +{4\over 3}
\end{equation}
if there is no scalar field in the bulk. One 
needs $C^{(4)}/\abs{C^{(3)}}\gg 1$ for the large 
gauge hierarchy, but this requires negative $k$ (or 
equivalently, positive $C^{(3)}$), for which the Higgs 
mass squared becomes negative, as seen 
from Eq. (\ref{shiki42}). Therefore, one cannot realize 
the large gauge hierarchy with only the fermions in the
bulk. One can say that in the model I the scalar
fields in the bulk are essential to realize the 
large gauge hierarchy. If we require the 
stability of the nontrivial vacuum, that is, the 
negative $C^{(3)}$, we have an inverse hierarchy
$M_W \gg M_c$.
%
%, which is out of our interest 
%in the paper. 
%
As for the model II, the flavor 
number of the massless bulk matter has no effect 
on the gauge hierarchy. As long as the bulk mass 
takes the appropriate values, the large gauge hierarchy 
is achieved. We care about the stability of 
the vacuum configuration, $F\geq 0$. Since $l=-4N_{adj}^{(-)}$ 
for the present case, $l$ is negative, so that $k~(\mbox{or~
equivalently},~C^{(3)})$ must be negative (positive). One 
easily finds a possible flavor set 
for $(k, l, m)=(-5, -4, -1)$, as an example, 
\begin{equation}
\left(N_{adj}^{(+)}, N_{fd}^{(+)}, N_{adj}^{(-)}, N_{fd}^{(-)}\right)
=\left(1, 3, 1, 0 \right),
\end{equation}
and the Higgs mass is obtained as 
\begin{equation}
m_H\simeq \left\{\begin{array}{cccc}
%141.1 & \mbox{GeV} & \mbox{for}& p=19,\\
%136.3 &\mbox{GeV} & \mbox{for} &p=18,\\
132.1 &\mbox{GeV} & \mbox{for} & p=17,\\
127.6 & \mbox{GeV} & \mbox{for}& p=16,\\
112.7 & \mbox{GeV} & \mbox{for}& p=13,\\
107.3 & \mbox{GeV} & \mbox{for}& p=12,\\
101.6 & \mbox{GeV} & \mbox{for}& p=11.
\end{array}
\right.
\end{equation}
Hence, the large gauge hierarchy and the Higgs 
mass are compatible in the model II even though 
the massless bulk matter is given by the 
fermions alone.
\par
%%%%%%%%%%%%%  
Let us discuss an important point for the 
scenario we have considered --- whether or not the 
condition $C^{(2)}=0$ is stable against 
higher loop corrections. The 
condition $C^{(2)}=0$ is realized by choosing 
the appropriate set of the flavor number at the 
one-loop level. If we have nonzero finite 
corrections at two (higher) loop level, our 
scenario considered in this paper no longer 
holds. In general, it may be natural to expect 
that we have nonzero finite corrections to 
the mass term at higher-loop level, but, at 
the moment, it may be too hasty to exclude 
the possibility of $C^{(2)}=0$ at the 
two (higher)-loop level.
\par
%%%%%%%%%
Recently, two loop calculation has been carried 
out in the five dimensional QED compactified 
on $M^4\times S^1$ \cite{maruyama}. In the 
paper, it has been reported that there is 
no finite correction to the Higgs mass from 
the two-loop level even though there is no 
concrete discussion to understand why it 
is so. As long as we have an example of the 
vanishing finite correction at the two-loop 
level, it does not seem unreasonable to expect 
the stability of 
%it may be allowed to believe the 
%stability of 
the condition against the higher loop corrections.
\par
%%%%%%%%%
In connection with the above discussion, it may be 
worth mentioning that there are examples, in 
which the loop correction is exhausted at 
the one-loop level (without supersymmetry). They 
are the coefficient of the axial 
anomaly \cite{adler} and the Chern-Simons 
coupling \cite{sakamoto}. As for the latter 
case, a simple reason for the two (higher) loop 
correction not to be generated comes from the invariance 
of the action under the large gauge transformation. 
Since the shift symmetry of the Higgs
potential can be regarded as the invariance under the large gauge
transformation, one may be able to 
%%%%%%%
%the two (higher)-loop correction to the 
%Chern-Simons action is believed not to be 
%generated due to the invariance of the action 
%under the large gauge transformation, which is 
%crucial for the argument. 
%%%%%%%%%%
prove that there is no two (higher)-loop correction
to the mass term of the Higgs potential.
%%%
%due to the invariance under the shift symmetry, which 
%is a sort of the large gauge transformation. 
%%%%%
In order to confirm it, we need more studies of 
the higher loop corrections 
to the Higgs potential (mass) in the
gauge-Higgs unification \cite{hoso}.
\par
%%%%%%%
We have considered the massless bulk matter 
belonging to the adjoint and fundamental 
representation under the gauge group $SU(3)$. One can
consider the higher dimensional representation such 
as ${\bf 10}, {\bf 15}$ of $SU(3)$ \cite{csaki}. In 
fact, these higher dimensional fields have been 
known to play an important role to enhance 
the Higgs mass and to obtain the large top 
Yukawa coupling for constructing realistic 
models. Therefore, it is interesting to study 
the effect of the higher dimensional field on 
the large gauge hierarchy in the scenario of 
the gauge-Higgs unification.   
\par
%%%%%%%%%%
Finally, let us comment on the possibility to realize 
the large gauge hierarchy if we have a term like 
\begin{equation}
B(\pi a)^2~{\rm ln}~(\pi a)^2,
\end{equation}
where $B$ is a constant. Since 
we are interested in the very small values of $a$, the Higgs 
potential is approximately given by
\begin{equation}
V= A(\pi a)^2 + B(\pi a)^2~{\rm ln}~(\pi a)^2
+O\left((\pi a)^4, (\pi a)^4{\rm ln}(\pi a)^2\right).
\label{shiki81}
\end{equation}
Then, the VEV is obtained in the desirable form as  
\begin{equation}
\pi a_0=\exp\left(-{A\over 2B}-\half\right),
\end{equation}
and if $A/B \gg 1 $, the large gauge hierarchy is realized. This is also
an interesting possibility, but unfortunately, we do not yet
understand the origin of the second term in Eq. (\ref{shiki81}). 
\par
%%%%%%%%%
Let us finally make a brief comment on the fermion
masses in the models we have studied. The mass scale for the
matter in the models is of the order of the weak scale or the
compactification scale $M_c$. Hence, it is difficult to reproduce the
realistic fermion mass spectrum in the models. This is the
common problem of the gauge-Higgs unification. Some attempts
have been done to overcome this problem 
in \cite{csaki}\cite{roma1}\cite{serone}.
%%%%%%%%%%
%\newpage
%%%%%%%%%%%%%%%%%%%
\begin{center}
{\bf Acknowledgments}
\end{center}  
This work was supported in part by a Grant-in-Aid for Scientific
Research (No. 18540275) by the Japanese Ministry of Education,
Science, Sports and Culture. M.S. would like to thank the 
Professor C. S. Lim for valuable discussions. K.T. is 
supported by the 21st Century COE Program at Tohoku University, and he would
like to express his cordial gratitude to his colleagues of Tohoku
University for discussions.
%%%%%%%%%%%%%%%%%%%%%%%%%%%%%
\vspace*{1cm}
%%%%%%%%%%%
%\begin{center}
%{\bf Appendix}
%\end{center}  
%%%%%%%%
%%%%%%%%%%%%% BIBLIOGRAPHY %%%%%%%%%%%%%%%%%%%%

%%%%%%%%%%%%%%%%%%%%%%%%%%%%
\end{document}